\documentclass[lettersize,journal]{IEEEtran}
\usepackage{amsmath,amsfonts}
\usepackage{algorithmic}
\usepackage{algorithm}
\usepackage{array}
\usepackage[caption=false,font=normalsize,labelfont=sf,textfont=sf]{subfig}
\usepackage{textcomp}
\usepackage{stfloats}
\usepackage{url}
\usepackage{verbatim}
\usepackage{graphicx}
\usepackage{cite}
\usepackage[center]{caption}
\usepackage{graphicx}
\usepackage{dcolumn}
\usepackage{bm}
\usepackage{siunitx}
\usepackage[colorlinks=true]{hyperref}
\hypersetup{
    colorlinks = true,
    linkcolor = blue,
    citecolor = blue,
    urlcolor=blue,
}
\begin{document}

\title{Machine learning assisted analysis of visible spectroscopy in pulsed-power-driven plasmas}



\author{Rishabh Datta, Faez Ahmed, and Jack D. Hare
\thanks{R. Datta and J. D. Hare are with the Plasma Science and Fusion Center, Massachusetts Institute of Technology, Cambridge,
USA}
\thanks{F. Ahmed is with the Department of Mechanical Engineering, Massachusetts Institute of Technology, Cambridge,
USA}
}



\maketitle

\begin{abstract}
We use machine learning models to predict ion density and electron temperature from visible emission spectra, in a high energy density pulsed-power-driven aluminum plasma, generated by an exploding wire array. Radiation transport simulations, which use spectral emissivity and opacity values generated using the collisional-radiative code PrismSPECT, are used to determine the spectral intensity generated by the plasma along the spectrometer's line of sight. The spectra exhibit Al-II and Al-III lines, whose line ratios and line widths vary with the density and temperature of the plasma. These calculations provide a 2500-size synthetic dataset of 400-dimensional intensity spectra, which is used to train and compare the performance of multiple machine learning models on a 3-variable regression task. The AutoGluon model performs best, with an R2-score of roughly $98\%$ for density and temperature predictions.  Simpler models (random forest, k-nearest neighbor, and deep neural network) also exhibit high R2-scores ($>90\%$) for density and temperature predictions. These results demonstrate the potential of machine learning in providing rapid or real-time analysis of emission spectroscopy data in pulsed-power-driven plasmas.
\end{abstract}


%
\IEEEpeerreviewmaketitle

\section{Introduction}

Spectroscopy is a powerful technique for inferring plasma parameters from emitted electromagnetic radiation. For instance, line widths and line ratios can be used to determine electron density and temperature \cite{griem2005principles,fantz2006basics,zhu2010optical}, velocity can be determined from the Doppler shift of spectral lines \cite{griem2005principles,hutchinson2002principles}, and magnetic field strength can be inferred from the Zeeman splitting of line radiation\cite{rochau2010applied,gomez2014magnetic}. The wide applicability of spectroscopy makes it an attractive tool for implementation in a variety of laboratory plasmas \cite{griem1992plasma,hutchinson2002principles,griem2005principles,sudkewer1981spectroscopic,astic1991visible}.

 In emission spectroscopy, a typical intensity spectrum can contain several peaks (called emission lines) overlaid on a continuum \cite{griem2005principles}. The lines correspond to bound-bound electron transitions in the ions of the plasma, while the continuum emission results from free-free (Bremsstrahlung emission) and free-bound electron transitions (recombination radiation) \cite{griem2005principles,hutchinson2002principles}. Line radiation generated by the plasma arises due to either collisional or radiative processes \cite{griem2005principles}. Collisional processes, such as electron impact excitation/de-excitation and three-body recombination, change the energy levels of bound electrons via collisions with other electrons \cite{griem2005principles,hutchinson2002principles}. Similarly, radiative processes, such as photoexcitation/de-excitation, induce energy transitions due to the interaction of bound electrons with photons \cite{griem2005principles,hutchinson2002principles}. Collisonal-radiative models balance the rates of excitation (and ionization) against that of de-excitation (and deionization), to determine the spectral emissivity and opacity of radiation emitted from the plasma \cite{fantz2006basics}.
 
 The typical approach to determining the ion density $n_i$ from emission spectra is to identify lines dominated by Stark (collisional) broadening, and then to compare the line widths with tabulated data \cite{griem2005principles,knudtson1987uv,astic1991visible}, or with the predictions of CR codes, such as PrismSPECT \cite{bailey2008diagnosis,macfarlane2013simulation}. Similarly, for the characterization of electron temperature $T_e$, we typically compare the intensity ratios of two or more lines (typically, inter-stage lines for which density changes have a small effect) with the predictions of CR models \cite{griem2005principles, knudtson1987uv,zhu2010optical}.

When the plasma is not optically thin, radiation transport, which describes how the energy distribution of radiation changes as it propagates through an absorbing, emitting, and/or scattering medium, must be adequately modeled for accurate interpretation of emission spectra. The optical thickness of a material to radiation of frequency $\omega$ is characterized by $\tau \equiv \int \alpha(\omega,s) ds$, which is the line-integral of the spectral opacity $\alpha(\omega)$ along the path $s$ \cite{drake2010high,hutchinson2002principles}. When $\tau \ll 1$, the plasma is optically thin, and the output spectrum is simply the line-integrated emissivity $\epsilon(\omega)$ of the plasma along the path $s$. Similarly, for $\tau \gg 1$, the plasma is optically thick, and the output spectrum (for a plasma in local thermodynamic equilibrium) is Planckian \cite{drake2010high,hutchinson2002principles}. In plasmas that are not optically thin, the radiation spectrum recorded by the spectrometer is significantly altered by radiation transport. Furthermore, if the plasma exhibits spatial inhomogeneity along the line-of-sight (LOS), the resulting spectrum may be dominated by strongly-emitting or absorbing regions.

In high-energy-density pulsed-power-driven systems, the condition for optical thinness may not be satisfied \cite{lebedev2019exploring, bell2009optical}. In this paper, our focus is to diagnose pulsed-power-driven plasmas for laboratory astrophysics applications, which typically exhibit ion densities and electron temperatures between $n_i \approx 1 \times 10^{17} - 1 \times 10^{19} \SI{}{\per \centi \meter \cubed}$ and $T_e \approx 1 - 50$ eV respectively \cite{lebedev2019exploring,datta2022structure,russell2022perpendicular,datta2023plasma}. Pulsed-power devices generate plasma flows by driving large currents (1-30 MA) through thin $\sim 10-\SI{100}{\micro \meter}$ diameter wires. These plasmas are not optically thin to visible radiation, therefore radiation transport modeling becomes important for spectral analysis. 

A key drawback of the aforementioned emission spectroscopy analysis approach is that it requires significant CR and radiation transport modeling for the analysis of a given spectrum. The use of machine learning (ML) models can reduce the computational time required for spectral analysis, especially for large batches of spectral data, and provide rapid real-time results during experimentation. Spectroscopy has previously been combined with supervised ML techniques, primarily in lower-density plasmas. Visible emission spectroscopy combined with regression methods and neural networks has been used to predict density and temperature in low-temperature low-density ($n_e \sim 1 \times 10^{10} \SI{}{\per \centi \meter \cubed}, \, T_e \sim 1$ eV)  laboratory plasmas \cite{park2021machine}. Similarly, neural networks have been used to predict electron energy distribution in low-temperature non-thermal plasmas \cite{mansouri2022methane}, and classifiers have been used for trace element and impurity detection in RF-generated plasmas \cite{shojaei2021application}. Neural network regressors and classifiers have also been shown to accurately predict electron temperature and divertor detachment from UV/XUV spectroscopy measurements in magnetic confinement fusion devices \cite{clayton2013electron,lee2016development,samuell2021measuring}. In the examples above, the labeled dataset for training is generated by simultaneous spectroscopy and independent density and temperature measurements with other diagnostics, like Thompson scattering or Langmuir probes \cite{clayton2013electron,lee2016development,samuell2021measuring}.  This eliminates the need for complicated theoretical or computational modeling. However, simultaneous independent measurements using secondary diagnostics are not always possible; for example, in experiments with poor diagnostic access or diagnostic unavailability. Moreover, alternative diagnostics such as Langmuir probes also perturb the plasma. Lastly, in experiments with low repetition rates, such as in pulsed-power plasmas, it is challenging to generate a purely experimental dataset for the data-hungry supervised ML task. In such situations, we must therefore rely on synthetic data for training. 

In this paper, we use a machine learning approach to predict electron density and temperature profiles along the spectrometer line of sight (LOS) from visible emission spectroscopy data in a pulsed-power-driven aluminum plasma. In contrast to previous work \cite{samuell2021measuring,park2021machine,lee2016development,mansouri2022methane}, which focuses on lower-density and/or optically thin plasmas, here we aim to characterize high-density non-optically thin spatially-inhomogeneous plasmas, characteristic of pulsed-power-driven plasmas generated using exploding wire arrays \cite{datta2022structure,russell2022perpendicular}.  We approach the problem in two parallel ways. In the first approach, we frame the prediction problem as a single-objective optimization problem, where we minimize the deviation between a simulated spectrum and a target spectrum. We generate the simulated spectra using CR calculations performed using PrismSPECT, which are then used to solve radiation transport in a spatially-inhomogeneous plasma. In the second approach, we solve a multi-variable regression problem, where we predict ion density $n_i(s)$ and electron temperature $T_e(s)$ as a function of position $s$ along the spectrometer LOS from a given spectrum. We train multiple supervised ML models --- linear regressor, k-nearest neighbor, decision trees, random forest, deep and convolution neural networks, and AutoGluon --- using synthetic data, generated from radiation transport simulations. The AutoGluon model performs best, with an R2-score of roughly $98\%$ for density and temperature predictions, and significantly reduces the computation time over optimization-based curve fitting methods. Our results demonstrate the potential of machine learning methods in providing rapid or real-time analysis of emission spectroscopy data in pulsed-power-driven plasmas.

\begin{figure}[t!]
\includegraphics[page=1,width=0.5\textwidth]{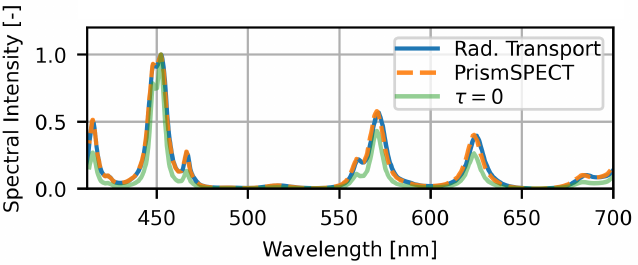}
\centering
\caption{Comparison of the output intensity calculated
by the radiation transport solver with that from a planar plasma simulation in PrismSPECT ($n_i = \SI{5e17}{\per \centi \meter \cubed}$, $T_e = 2.5$ eV, and $L = \SI{10}{\milli \meter}$), and with a zero opacity $\tau = 0$ case.} 
\label{fig:rad_tran_comparison}
\end{figure}
 
\section{Dataset generation}

\subsection{Radiation Transport Modeling}
\label{sec:rad_tran}

We use PrismSPECT \cite{macfarlane2013simulation} to compute emissivity $\epsilon (\omega)$ and opacity $\alpha (\omega)$ values for an aluminum plasma, in the visible range of the electromagnetic spectrum ($\SI{400}{\nano \meter} < \lambda \, < \SI{700}{\nano \meter}$). We use a steady-state non local thermodynamic equilibrium model with Maxwellian free electrons. We run 10,000 PrismSPECT simulations, for electron temperatures linearly distributed in the range $T_e \subseteq [0.5,25]$ eV, and ion density logarithmically distributed between $n_i \subseteq [1 \times 10^{16}, 1 \times 10^{19}] \, \SI{}{\per \centi \meter \cubed}$. 

 Our in-house radiation transport solver computes the output intensity spectrum $I_\omega(s)$ given spatially-varying emissivity $\epsilon_\omega(s)$ and opacity $\alpha_\omega(s)$ values, by solving the steady-state radiation transport equation along the 1D path $s$ \cite{drake2010high}:
\begin{equation}
 \partial_s{I_\omega(s)} = \epsilon_\omega(s) - \alpha_\omega(s) I_\omega(s)
 \label{eq:radtrans}
\end{equation}

\begin{figure}[b!]
\includegraphics[page=4,width=0.5\textwidth]{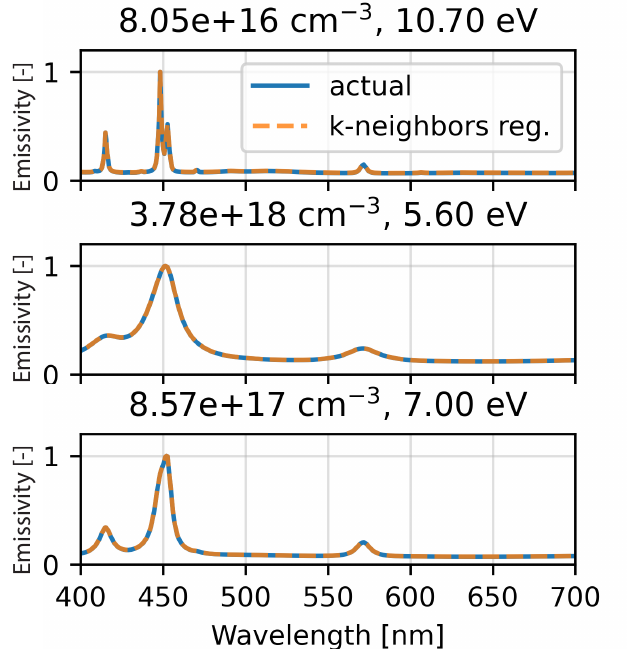}
\centering
\caption{Comparison of emissivity predictions made by a KNN model trained on PrismSPECT simulation results. We compare results for randomly-chosen members of the test set. Emissivity is scaled to [0,1].}
\label{fig:KNN}
\end{figure}

Emissivity and opacity values vary along the LOS due to spatial variations in density and temperature. PrismSPECT computes $\epsilon_\omega$ and $\alpha_\omega$ for spatially-homogeneous plasmas. To construct the spatially-varying emissivity and opacity, we first assume some m-dimensional density and temperature distributions $n_i(s), \, T_e(s) \in \mathcal{R}^m$ along the LOS. Here, $m$ is the number of points required to capture the spatial variation along the LOS. We then calculate the emissivity and opacity values at each position $s$ along the LOS from the values of $n_i(s)$ and $T_e(s)$ at that location. The radiation transport solver then determines the output intensity distribution by solving \autoref{eq:radtrans}. We use a spectral resolution of $3.75 \times 10^{-3}$ eV for our radiation transport calculations, which results in a [$400 \times 1$] dimensional intensity spectrum for each data point. 

\begin{figure}[b!]
\includegraphics[page=2,width=0.5\textwidth]{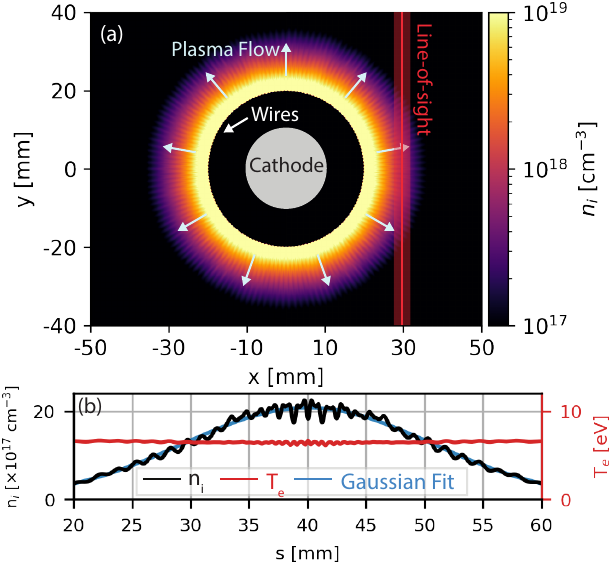}
\centering
\caption{(a) Simulated ion density at peak current, generated by a $\SI{40}{\milli \meter}$ diameter exploding wire array with 150 aluminum wires, driven by a 10 MA current pulse ($\SI{300}{\nano \second}$ rise time). This simulation was performed using GORGON, a two-temperature resistive MHD code (b) Variation of density and temperature along a chordal line-of-sight (LOS) as shown in (a).}
\label{fig:gorgon}
\end{figure}

In \autoref{fig:rad_tran_comparison}, we compare the output intensity generated by the radiation transport solver with that from a planar plasma simulation in PrismSPECT. Here, the ion density and electron temperature of the plasma are $n_i = \SI{5e17}{\per \centi \meter \cubed}$ and $T_e = 2.5$ eV respectively, while the size of the plasma is $\SI{10}{\milli \meter}$. The planar plasma simulation includes the effect of radiation transport, for the case of a homogenous (i.e. constant density and temperature) slab of a specified size. The output of the radiation transport solver agrees with that from PrismSPECT. \autoref{fig:rad_tran_comparison} also shows the case for which the optical thickness is set to zero. For the $\tau = 0$ case, lines with high opacities are no longer damped by absorption from the plasma, and thus, the line ratios are significantly modified when compared to the case with radiation transport. This illustrates that in our pulsed-power-driven plasma of interest, optical thickness is important, and must be included in the spectroscopy analysis. 

The radiation transport solver requires emissivity and opacity calculated at each density and temperature value along the LOS as inputs to generate the intensity spectrum. We interpolate the emissivities and opacities for the intermediate values not simulated with PrismSPECT, using a k-nearest neighbor (KNN) regressor, trained on the output of the 10,000 PrismSPECT simulations. To evaluate the performance of the regressor, we compare the predicted spectra with the previously unseen emissivity spectra in the test set. As observed in \autoref{fig:KNN}, where we compare the predicted emissivity with the actual emissivity for randomly-chosen members of the test set, the predictions agree well with the actual spectra. The coefficient of determination (also called the R2-score) is a commonly-used metric to characterize the performance of regression, and is defined as:
\begin{equation}
    R^2=1-\frac{\sum\left(y_i-\hat{y}_i\right)^2}{\sum\left(y_i-\bar{y}\right)^2}
    \label{eq:R2}
\end{equation}
Here, $y_i$ is the predicted value, $\hat{y}_i$ is the actual value, and $\bar{y}\equiv1/n\sum_{i=0}^n y_i$ is the mean of the actual values. For the problem above, the KNN regressor exhibits an R2-score of 99.62\%, showing that the model accurately reproduces the emissivity and opacity spectra for density and temperature values not included on the simulation grid in PrismSPECT.

\subsection{Density and Temperature in Exploding Wire Arrays}

Exploding wire arrays, which consist of a cylindrical cage of wires around a central cathode, are commonly used sources of pulsed-power-driven plasma for laboratory astrophysics experiments \cite{lebedev2019exploring,datta2022structure,russell2022perpendicular}. The magnetic field is oriented in the azimuthal direction inside the wires, and results in a $\bf{j \times B}$ force that accelerates the ablating plasma radially outwards from the wires. Due to radially-diverging flows, the density decays rapidly with distance from the wires. This can be observed in \autoref{fig:gorgon}a, which shows the simulated ion density distribution generated by a $\SI{40}{\milli \meter}$ diameter exploding wire array with 150 aluminum wires, driven by a 10 MA current pulse with a $\SI{300}{\nano \second}$ rise time. This simulation was performed using GORGON --- a two-temperature Eulerian resistive magnetohydrodynamic (MHD) code \cite{ciardi2007evolution}.

As discussed before, we require $m$-dimensional arrays to fully capture the spatial variation in density and temperature along the  LOS. However, we can use simplifying assumptions to make the problem more computationally tractable. If the emission is recorded along a chordal LOS, as shown in \autoref{fig:gorgon}a, the density variation  can be approximated as Gaussian (see \autoref{fig:gorgon}b) i.e. $n_i(s) = n_0 \exp[ (s-s_0)/ 2 \sigma^2] $. Here, $n_0$ is the peak density, $\sigma$ is the standard deviation of the Gaussian function, and we set the mean $s_0$ to half the total path length. The density distribution shown in \autoref{fig:gorgon} is consistent that measured experimentally with laser imaging interferometry in previous pulsed-power experiments \cite{datta2022structure}. Our MHD simulations (\autoref{fig:gorgon}b) and previous experimental measurements also show little spatial variation in the temperature due to short thermal diffusion time in pulsed-power plasmas \cite{russell2022perpendicular}. Therefore, we approximate the temperature to be constant along the LOS i.e. $T_e \neq T_e(s) = T_0$. This allows us to reduce our $2 \times m$ dimensional problem to just 3 variables --- $n_0$, $T_0$, and $\sigma$. 
Both the density and temperature in \autoref{fig:gorgon}b also exhibit small amplitude modulations, which arise due to oblique shocks resulting from the azimuthal expansion of plasma from the discrete wires \cite{swadling2013oblique}. Our radiation transport calculations, however, show that the effect of these modulations on the recorded intensity spectrum is small.

\begin{figure}
\includegraphics[page=3,width=0.5\textwidth]{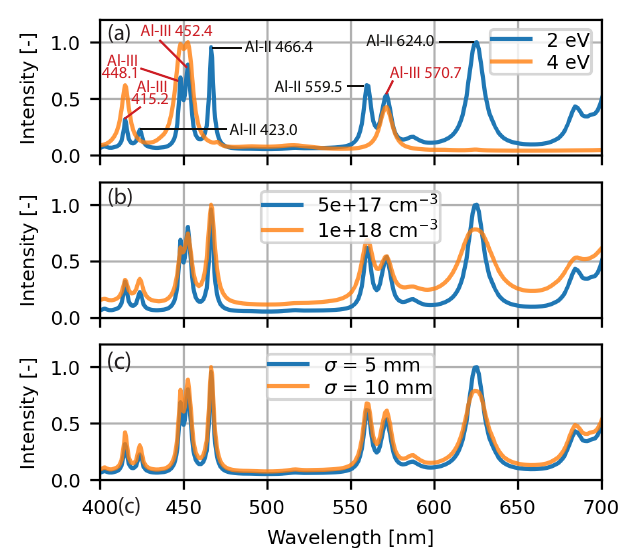}
\centering
\caption{Normalized spectral intensity simulated by the radiation transport solver. In (a)-(c), the blue curves correspond to output spectra generated for a Gaussian density variation ($n_0 =\SI{5e17}{\per \centi \meter \cubed}$, $\sigma = \SI{5}{\milli \meter}$), and constant temperature ($T_e = 2$ eV) along the spectroscopy LOS. (a) Change in the intensity spectrum with increasing temperature. (b) Change in the intensity spectrum with increasing density $n_0$. (c) Change in the intensity spectrum with increasing $\sigma$.}
\label{fig:examples}
\end{figure}

\autoref{fig:examples} shows a synthetic intensity spectrum, generated by the radiation transport solver, with values $n_0 =\SI{5e17}{\per \centi \meter \cubed}$, $T_e = 2$ eV, and $\sigma = \SI{5}{\milli \meter}$. Here, we normalize the spectrum between $[0,1]$ by dividing by the maximum intensity. The spectrum exhibits Al-II and Al-III lines, which correspond to transitions in singly- (Mg-like) and doubly-ionized (Na-like) aluminum respectively. When the temperature is increased (\autoref{fig:examples}a), the relative intensity of the Al-III lines compared to the Al-II lines increases. This is expected because the ionization is higher at a higher temperature, and thus, the relative population of the higher-$\bar{Z}$ Al-III ions increases relative to the singly-ionized Al-II ions. The Al-II and Al-III lines only appear simultaneously between 1.5-3.5 eV. In \autoref{fig:examples}a, at 4 eV, the Al-II lines are  completely suppressed. When we increase the density (\autoref{fig:examples}b), the lines not only become broader (due to Stark broadening) but the line ratios change as well. This is because increasing the density also increases the optical thickness, and the optically-thick lines are damped more strongly. This can be observed in \autoref{fig:examples}b when we compare the relative intensity of the Al-II \SI{466.4}{\nano \meter} line (which is relatively optically thin) with that of the higher-opacity Al-II \SI{624.0}{\nano \meter} line. The more optically-thick Al-II \SI{624.0}{\nano \meter} line is strongly damped at higher densities. Finally, changing the value of $\sigma$ (\autoref{fig:examples}c), also changes the line ratios because the optical thickness increases with the size of the plasma; however, the sensitivity of the spectrum to changes in $\sigma$ is relatively smaller than that in density and temperature. 

To generate our training dataset for the machine learning task, we randomly sample values of $n_0 \subseteq [0.5 \times 10^{17}, 1 \times 10^{19}] \SI{}{\per \centi \meter \cubed}$, $\sigma \subseteq [5,40]$ mm, and $T_0 \subseteq [0.5,25]$ eV from a uniform distribution. Our radiation transport solver then uses these sampled values to calculate intensity spectra for each $n_0$, $\sigma$, and $T_e$. We generate a total of 2500 [$400 \times 1$] intensity spectra to use as training data for our 3-variable regression problem. Lastly, we scale and normalize the values of $n_0 \, , \sigma$ and $T_0$ so that they lie within the interval $\subseteq [0,1]$. We use linear scaling for the temperature and $\sigma$, and logarithmic scaling for the density. We also scale the intensity output of the radiation transport solver to the range $\subseteq [0,1]$. This means we only use the shape of the intensity spectrum for our prediction, which obviates the need for absolute intensity calibration.

\section{Methodology}

\subsection{Single-Objective Optimization}

Our goal is to predict density and temperature profiles given a measured intensity spectrum $I_\text{target}$. One way to frame this problem is as a single-objective optimization problem:

\begin{equation}
\begin{split}
    \min _\mathbf{x} : f(\mathbf{x})&=\text{MSD}\left( I_{\text {target }}-I(\mathbf{x})\right) \\
    \text { where: } \mathbf{x}&=\left[n_0, \sigma, T_0\right] \\
    \text{s.t.} : 0.5 \times 10^{17} &\leq n_0 \, [\SI{}{\per \centi \meter \cubed}] \leq 1 \times 10^{19} \\
    \qquad  5 &\leq \sigma \, [\SI{}{\milli \meter}] \leq 40\\
    \qquad  0.5 &\leq T_0 \, [\text{eV}] \leq 25\\
\end{split}
\end{equation}

We minimize the mean squared deviation (MSD) between the target and predicted intensities. The objective function can be represented as:
\begin{equation}
\text{MSD} = \frac{1}{N}\sum_i^N[I_{pred}(\omega_i) - I_{target}(\omega_i)]^2
\label{eq:MSD}
\end{equation}

Here, $I_{\text{{pred}}}(\omega_i)$ and $I_{\text{{target}}}(\omega_i)$ are the simulated and target intensities at frequency $\omega$, and $N = 400$ is the size of the intensity spectrum. We use the radiation transport solver described in \autoref{sec:rad_tran} to generate the predicted intensities. 

We perform the optimization using a $(\mu + \lambda)$ genetic algorithm (GA) implemented using the pymoo package in Python. The GA optimization algorithm iteratively searches for solutions that minimize the objective function over multiple generations \cite{mirjalili2019genetic}. In each generation, the best-performing solutions are selected and included in the population for the next iteration. Solutions are combined in each iteration in a process called crossover to create offspring solutions. The solutions are also subject to random changes in the values of the variables ${\bf x}$ to increase the diversity of the solutions, in a process called mutation. For the optimization here, we use a randomly generated initial population size of 150, with simulated binary crossover and polynomial mutation (probability = 0.5, distribution index $\eta_c = 1$). We terminate the optimization when the MSD of the solutions becomes lower than a specified threshold. We repeat the optimization 50 times with different starting seeds to construct a family of optimal solutions. We exclude solutions from runs in which the GA gets stuck in a local minimum, where the objective function does not converge to a value below the required threshold.

\subsection{Multi-Variable Regression}

\label{sec:MLmodels}

An alternative approach is to formulate the problem as a multi-variable regression problem. Given an input of an unseen intensity spectrum $I_\text{target}$, we predict the corresponding (normalized) values of $n_0^*$,  $T_0^*$, and $\sigma^*$ using machine learning-based regression models. Here, we compare the performance of multiple regressors for our three-variable regression task --- linear regression (LR), k-nearest neighbor regressor (KNN), decision trees (DT), random forest (RF), deep neural network (DNN), and a 1D convolution neural network (1D-CNN). The choice of regression algorithm often represents a trade-off between model precision and interpretability. Simpler models such as linear regression and KNN models are relatively easier to understand and interpret, whereas deep neural networks and AutoML, which often provide high performance, require more training time, and are challenging to interpret \cite{james2013introduction}.

The linear regression, KNN, decision trees, gaussian process regressor, and random forest  models are implemented using the scikit-learn \cite{pedregosa2011scikit} package in Python. The KNN algorithm predicts values based on the distance from the $k$ nearest data points in the training set \cite{james2013introduction}. Here, we use $k=8$ and Euclidian distance for our KNN regressor. Decision trees follow a flowchart-like structure, and make predictions by asking questions at each level \cite{james2013introduction}. We use a DT regressor with a  depth of 5 and a minimum sample split of 5. Random forests are ensembles of many decision trees \cite{james2013introduction}. Our random forest regressor uses 140 estimators; minimum samples required for a split is 4, and the minimum samples per leaf are 5. 

We use the TensorFlow \cite{abadi2016tensorflow} package to construct a 3-layer deep neural network (DNN). Neural networks consist of multiple `hidden' layers, consisting of several neurons, sandwiched between an input and an output layer \cite{anderson1995introduction}. In our DNN architecture, each layer is a fully-connected dense layer with 100 neurons (i.e. each neuron is connected to every other neuron in the previous and next layers), with ReLu activation functions, and padded with a batch normalization layer and a dropout layer ($p = 0.3$) to help prevent overfitting. The final layer consists of a 3-dimensional dense layer. Similarly, the 1D-CNN consists of two 1D convolution layers (kernel size = 3) with filter sizes of 8 and 16 respectively, and leaky ReLU activation layers. The convolution layer performs a convolution operation using the specified kernel on the input from the previous layer \cite{anderson1995introduction}. Each convolution layer is padded with a batch normalization layer, and a $p = 0.1$ dropout layer.  The convolution layers are followed by a 200-neuron dense fully-connected layer, and a 3-dimensional output layer. Both neural networks use the Adam optimizer with a $\SI{0.5e-4}{}$ learning rate, and a mean squared error (MSE) loss function. The key difference between the DNN and CNN architectures is that the DNN treats the input vector as a 400-dimensional vector of parameters, whereas the CNN treats it as a 1D image, and therefore, has information about the relative spectral location of each element in the input vector. 

\begin{figure}[b!]
\includegraphics[page=5,width=0.48\textwidth]{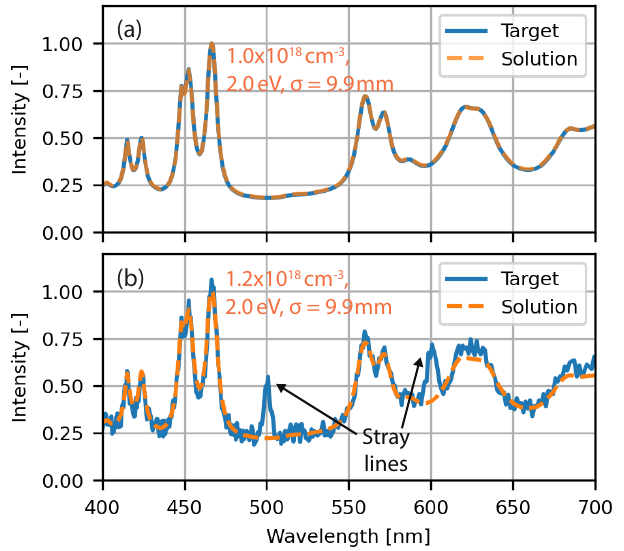}
\centering
\caption{(a) Comparison of target intensity spectra with spectra determined via optimization for the test case $n_0 = \SI{1e18}{\per \centi \meter \cubed}, \, T_0 = 2 \, \text{eV}, \, \sigma = \SI{10}{\milli \meter}$. The target spectrum is the solid blue curve, while the predicted spectrum and the predicted values are  in orange. (b) Robustness of the fitting to noise and stray lines.}
\label{fig:optimization}
\end{figure}

Finally, we also implement a tabular AutoGluon model using Python's AutoGluon package \cite{erickson2020AutoGluon}. AutoGluon provides an automated approach to machine learning, by automatically comparing and combining the performance of many different models. The performance of AutoGluon has previously been shown to exceed that of more traditional ML models \cite{erickson2020AutoGluon}.

Each regressor is trained on the $2500 \times 400$ synthetic spectra with a 2000:500 split between the training and test sets, and $k=3$ stratified k-fold cross-validation for the training set. Members of the test set are not shown to the ML model during the regression task, and are used to evaluate the regression performance after training.  The hyperparameters described in this section were determined using hyperparameter optimization implemented using the Python package Optuna \cite{akiba2019optuna}. 

\begin{table*}\centering
\caption{\centering Comparison of the performance of different ML models on the regression task}
\begin{tabular}{lccc c}
\hline
Model & Training Time (s) & R2-Score($\%$) & 2 Var. R2-Score($\%$) & Median MSD [$\times 10^{-4}$] \\
\hline
Linear Regression (LR)  & 0.3 & $-41.54 \pm 80.76$ & $64.00 \pm 10.2$ & 1.1 \\
Decision Tree (DT)  & 0.35 & $53.22 \pm 4.5$ & $89.9 \pm 2.57$   & 5.3\\
k-Nearest Neighbor & 0.02 & $67.89 \pm 2.8$ & $96.25 \pm 1.8$ & 0.4  \\
Random Forest (RF) & 6.41 & $ 67.06 \pm 1.93$ & $96.58 \pm 0.5$ & 0.4  \\
Deep Neural Net. (DNN) & 240.12 & $71.54 \pm 1.2$ & $94.52 \pm 1.1$  & 4.2\\
1D Conv. Neural Net. (1D-CNN) & 260.17 & $65.34 \pm 3.2$ & $91.69 \pm 1.4$  & 5.7\\
{\bf AutoGluon} & {\bf 4760.11} & $\mathbf{74.20 \pm 2.1}$ & $\mathbf{98.98 \pm 1.4}$ & $\mathbf{0.3}$ \\
\hline
\end{tabular}
\label{tab:table}
\end{table*}

We use two metrics to characterize the performance of the different regression models. These are the coefficient of determination (also called the R2-score), and the mean squared deviation (MSD) from the simulated curve. The R2-score is defined in \autoref{eq:R2}, and is calculated from the predicted and actual values of $n_0^*, T_0^*$ and $\sigma^*$. Similarly, the MSD (\autoref{eq:MSD}) measures the deviation of the predicted intensity spectrum from the actual spectrum. Here, the predicted spectrum is determined by feeding the predicted values of $n_0^*$, \, $T_0^*$ and $\sigma^*$ from the ML model into the radiation transport solver. The R2-score and MSD are calculated for the $N = 500$ test set, which contains spectra not previously seen by our ML models. For good performance, we aim to maximize the R2-scores, and minimize the MSD. The R2-score may provide spurious performance metrics in case of non-unique solutions, whereas, the MSD characterizes how close the predicted spectra are to the actual spectra, allowing us to overcome this issue. 

\section{Results and Discussion}

\subsection{Optimization results}
\label{sec:opt_results}

Using GA-based optimization, we predict values of $n_0,\, T_0$ and $\sigma$ for several test cases. \autoref{fig:optimization}a compares the target intensity spectra with spectra determined via optimization for a randomly-selected test case ($n_0 = \SI{1e18}{\per \centi \meter \cubed}, \, T_0 = 2 \, \text{eV}, \, \sigma = \SI{10}{\milli \meter}$). The predicted values of $n_0,\, T_0$ and $\sigma$ reproduce the target spectrum well. For the test case shown in \autoref{fig:optimization}a, the MSD from the target spectrum is roughly $\SI{6e-5}{}$, and the predicted values are $n_0 = \SI{1.0e18}{\per \centi \meter \cubed} \pm 7\%$, $T_0 = 2.0 \, \text{eV} \pm 1\%$, and $\sigma = \SI{9.9}{\milli \meter} \pm 20\%$. The range in the value of $\sigma$ for the family of optimal solutions is comparatively higher than that in the density and temperature. This is consistent with our observation in \autoref{fig:examples}c, which shows that changes in $\sigma$ generate linear changes in the optical depth, whereas those in $n_0$ and $T_0$ result in larger non-linear changes in opacity and relative intensities.

In many real situations, the experimental data that we want to fit may be noisier than the synthetic spectrum shown in \autoref{fig:optimization}a. Furthermore, the spectrum may also be contaminated by stray lines, generated by impurities in the plasma, or by radiation emitted from other photoionized surfaces. In order to test the robustness of the prediction, we add noise and stray lines to the synthetic target spectrum (see \autoref{fig:optimization}b). The predicted solution reproduces the target spectrum well, despite the added noise and stray radiation. In this case, the MSD of the optimum solutions is about $\SI{3e-3}{}$, which as expected, is larger than that for the smooth case. The predicted solutions are  $n_0 = \SI{1.2e18}{\per \centi \meter \cubed} \pm 20\%$, $T_0 = 2.0 \, \text{eV} \pm 2\%$, and $\sigma = \SI{9.9}{\milli \meter} \pm 30\%$. There is a larger uncertainty in the solutions for the noisy target spectrum when compared to the smooth target spectrum (\autoref{fig:optimization}a); however, the predicted solutions still include the values used to generate the test case.

Although optimization using  GA provides good results, the computational time is high ($>10$ min per prediction). This is primarily because the GA has to simulate a large number of potential candidates iteratively using our radiation transport model over several generations before convergence is achieved. Although we use GA as an example here, other optimization-based curve-fitting algorithms can also exhibit high computation times. This method can be highly effective with small data sets; however, it can be less attractive in cases where we must analyze large datasets of spectra, or when we require quick or real-time analysis of spectral information. The use of ML-based regression models, discussed in the next section, can be more useful in such situations.

\begin{figure}[t!]
\includegraphics[page=8,width=0.48\textwidth]{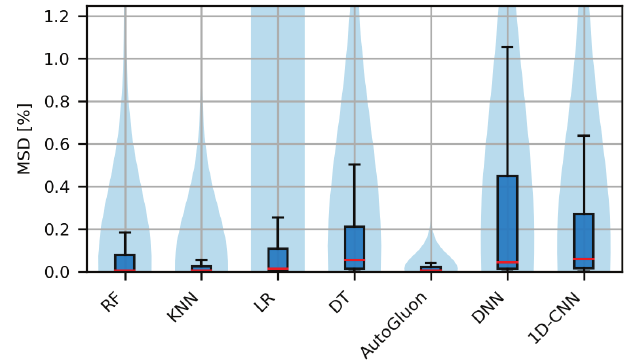}
\centering
\caption{Violin plots of the MSD of the predicted intensity spectra from the actual intensity spectra in the test set. The red lines represent the medians of the distribution, the blue rectangle represents the inter-quartile range, and the end-caps represent $1.5\times$ the inter-quartile range. The blue-shaded regions show the shape of the full distribution.}
\label{fig:MSD}
\end{figure}

\begin{figure*}
\includegraphics[page=6,width=1\textwidth]{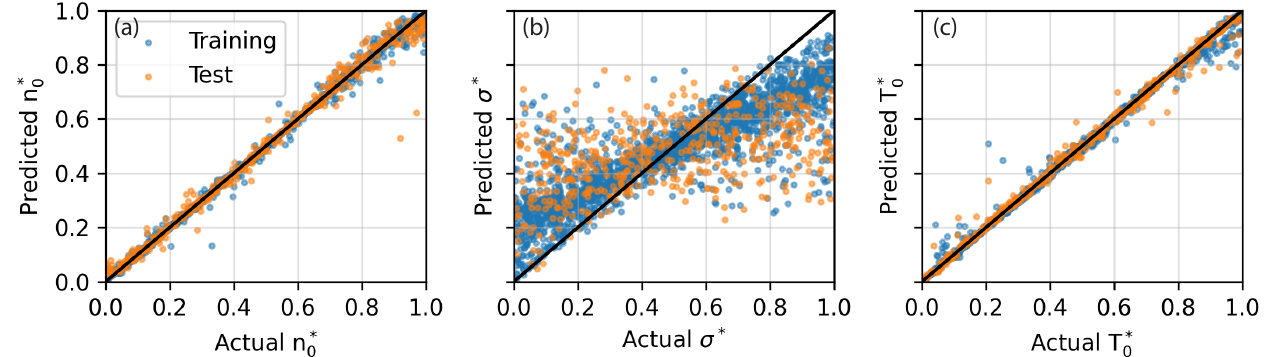}
\centering
\caption{Distribution of predicted values of $n_0^*$, $T_0^*$ and $\sigma^*$ against the actual values for the AutoGluon Model. Here, the values of peak density, temperature, and the spread of the density function $\sigma$ have been normalized. Blue points compare the values for the training set, while orange points do so for the test set.}
\label{fig:auto}
\end{figure*}

\subsection{Multi-variable regression using machine learning}

The training of ML models can be computationally time-intensive; however, once the training is complete, large datasets can be evaluated rapidly. The training time of models typically depends on the complexity of the model. \autoref{tab:table} compares the performance of the ML regressors described in \autoref{sec:MLmodels}. Here, the R2-score is computed using the predicted and actual values of all three variables, while we compute the two-variable R2-score using only the value of density $n_0^*$ and temperature $T_0^*$. AutoGluon was the best-performing model, with an R2-score of $74.20\% \pm 2.1 \%$. Since AutoGluon trains and compares the performance of multiple different models simultaneously, the training time was high compared to the other models. The next best-performing models (deep neural network, Random Forest, and k-nearest neighbor) exhibited R2-scores between $67-71 \%$ and shorter training times. The two-variable R2 scores, calculated for density $n_0^*$ and  temperature $T_0^*$ only, are roughly $30 \%$ higher than the three-variable R2-scores, which shows that the models predict $n_0^*$ and $T_0^*$ with better accuracy than the third variable $\sigma^*$. There is a larger uncertainty in the prediction of $\sigma^*$, because as observed in \autoref{fig:examples}c and in \autoref{sec:opt_results}, changes in $\sigma^*$ generate linear changes in the optical depth, whereas those in $n_0^*$ result in larger non-linear changes. We also find that in cases where $\sigma^*$ is under-predicted, the density is slightly over-predicted and vice versa, indicating that the model relies on small changes in density rather than on $\sigma^*$ to incorporate the effects of changing optical depth.  The majority of models, with the exception of linear regression and decision trees, exhibit two-variable R2-scores $> 90 \%$, showing the effectiveness of these models in predicting $n_0^*$ and $T_0^*$, despite the relatively poorer prediction of $\sigma^*$. AutoGluon, once again, exhibits the highest two-variable R2-score ($98.98\% \pm 1.4 \%$).

In these models, we only use the shape of the intensity spectrum to make predictions. This can be advantageous as it obviates the need for absolute intensity calibration of the intensity spectra. However, adding information about the absolute intensity can potentially improve the predictions, as the absolute intensity also depends on the temperature, density, and size of the plasma. This would, however, require accurate modeling of intensity attenuation and losses in the optics used for the experiment. When we include the absolute intensity in the training data (that is, we do not normalize the intensity between $[0,1]$), the models show improved prediction of $\sigma$, and the R2-scores increase by roughly 3-7\%. 

\autoref{fig:MSD} compares the MSD of the predicted intensity spectra from the actual intensity spectra in the test set for the different ML models. Here, we use the predictions of $n_0, \, T_0 \, \text{and } \sigma$ from the ML regressors as inputs into the radiation transport solver to determine the predicted spectra. The red lines represent the medians of the distributions, the blue rectangles represent the interquartile range, and the shaded regions show the shape of the full distribution. As expected, the predictions of the AutoGluon model, which exhibit the highest R2-score, also exhibit the lowest MSD, i.e. the predictions match the original intensity spectra well. The median MSD is relatively small ($<0.1\%$) for all the models, however, for models with lower R2-scores, the distribution of the MSD exhibits a larger spread and extends to higher values. Interestingly, although the DNN exhibits R2-scores similar to the RF and KNN models, its MSD distribution is significantly wider. For the DNN, the distribution is also wider than for the 1D-CNN model, which exhibited lower R2-scores. This may indicate that although the predictions of $n_0, \, T_0 \, \text{and } \sigma$ made by the DNN are close to the actual values, the values are consistently off, resulting in a relatively larger deviation between the predicted intensity calculated from these values and the actual intensity used to predict the values.

To gain further insight into the performance of the models, we plot the distribution of predicted values against the actual values. For the best-performing AutoGluon model (\autoref{fig:auto}), we find that predicted values of $n_0^*$ and $T_0^*$ from the test set exhibit the least deviation from the diagonal, whereas predicted values of $\sigma^*$ exhibit a relatively larger deviation from the diagonal, as expected. However, AutoGluon is approximately able to capture the relative trend in the values of $\sigma^*$, as seen in the positive slope of the distribution in \autoref{fig:auto}b. This is in contrast to other models, which exhibit larger deviations in the predictions of $\sigma^*$. \autoref{fig:auto}a shows that the predicted values of density deviate from the actual values more significantly at very low ($n_0 < \SI{1e17}{\per \centi \meter \cubed}$) and very high densities ($n_0 >\SI{5e18}{\per \centi\meter \cubed}$). At low density, the effects of optical depth and Stark broadening are smaller, which leads to larger uncertainties in the prediction of the density. Our radiation transport simulations also show that at higher densities, line radiation is strongly damped due to the opacity effects, and continuum emission begins to dominate. This may explain why the predicted density deviates more significantly at these higher densities.  For the temperature predictions (\autoref{fig:auto}c), the deviations are relatively small for $T_0 < 3$ eV, and become more significant at higher temperatures. Below 3 eV, as seen in \autoref{fig:examples}a, the spectrum contains both Al-II and Al-III lines, and the relative intensities of the inter-stage lines are a strong function of temperature. However, at higher temperatures, only the Al-III lines are present in the spectrum, and continuum emission also begins to dominate, which makes it relatively harder to predict $T_0$.



\begin{figure}[b!]
\includegraphics[page=7,width=0.5\textwidth]{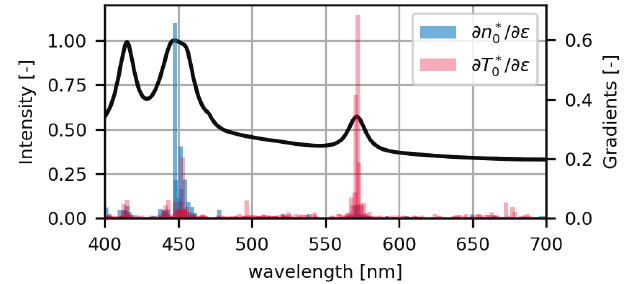}
\centering
\caption{Sensitivity of the predictions of density and temperature to variations in the spectral intensity value at different wavelengths for a randomly-chosen intensity spectrum from the test set. }
\label{fig:saliency}
\end{figure}

A key challenge with more complicated ML models, such as neural networks and AutoGluon, is the lack of interpretability. To gain insight into features that inform the prediction, we investigate the relative importance of different parts of the intensity spectrum for the prediction of density and temperature. In order to do so, we calculate the sensitivity of the density and temperature predictions to perturbations in the spectral intensity value at different wavelengths. \autoref{fig:saliency} shows the sensitivity map computed for the AutoGluon model for a randomly-selected member of the test set. For this intensity spectrum, the actual values of density, temperature, and $\sigma$ are $\SI{1.4e18}{\per \centi \meter \cubed}$, $7.3$ eV, and $\SI{25}{\milli \meter}$, and the predicted values are $\SI{1.5e18}{\per \centi \meter \cubed}$, $6.9$ eV, and $\SI{22}{\milli \meter}$.  We approximate the gradients in density and temperature ($\partial n_0^*/\partial\epsilon$ and $\partial T_0^*/\partial\epsilon$) by perturbing the input intensity spectrum at a given wavelength by a small value $\epsilon$, and dividing the change in the predicted value ($|\delta n_0^*|$ or $|\delta T_0^*|$) by the perturbation $\epsilon$. Here, we pick the perturbation $\epsilon$ from a Gaussian distribution of amplitude 0.1, and the mean gradients are calculated over multiple iterations to determine the final value. Such sensitivity maps are commonly used in image classification problems to identify parts of an image that may contribute to the final classification \cite{selvaraju2017grad}. Here, we use it to identify features of the spectra that contribute to the final predictions in our regression problem. In \autoref{fig:saliency}, as expected, the lines (Al-III 415.2 nm, Al-III 448.1 nm, Al-III 452.5 nm, and  Al-III 570.7 nm) contribute significantly to the prediction of density and temperature, whereas parts of the spectrum that correspond to the continuum are less important. This can be observed from the large gradients $\partial n_0^*/\partial\epsilon$ and $\partial T_0^*/\partial\epsilon$ at the positions of these lines. The Al-III 570.7 nm line appears to be particularly important for the temperature prediction, while the Al-III 448.1 nm and Al-III 452.5 nm lines (which have merged at this higher density) are relatively more important to the density prediction. In addition to the locations of the Al-III lines, smaller peaks in the gradients, particularly in temperature, also appear at parts of the spectrum devoid of lines. The locations of these peaks correspond to Al-II emission lines that appear predominantly at lower temperatures ($0.5-1.5$ eV), indicating that the absence of these lines contributes, although less significantly, to the temperature prediction.

Another limitation of these models is the reliance on synthetic data, which, in turn, is affected by the uncertainty and assumptions in the theoretical modeling used to generate the synthetic dataset. Benchmarking the results using independent diagnostics can be one way to probe the applicability of these ML models to real experimental data. However, as mentioned earlier, independent measurements are not always possible, and the uncertainty in CR modeling is inherent in the analysis of most spectroscopic data. As discussed in \autoref{sec:opt_results}, the experimental data can also be noisy, and include background radiation and contamination by impurities. When we use the trained ML models to make predictions from noisy spectra contaminated with stray radiation, the R2-scores typically fall by $10-20\%$.  For a given experiment, the instrument response, the bit depth of the spectrometer, and attenuation by the optics may also need to be properly included in the synthetic dataset. The analysis of experimental spectra using these methods will be pursued in a future publication.

\section{Conclusions}

We explore the use of machine learning (ML) methods for rapid spectroscopic analysis of emission spectra in the visible regime. Our goal is to predict density and temperature in a pulsed-power-driven aluminum plasma generated by an exploding wire array. In contrast to previous work, which has typically focused on low-density homogenous plasmas, we aim to diagnose a high energy density non-optically thin plasma, which necessitates the use of radiation transport calculations to accurately model the recorded intensity spectrum. These radiation transport calculations use spectral emissivity and opacity values computed using the collisional-radiative code PrismSPECT. Consistent with previous observations in exploding wire arrays, we assume a constant temperature and a Gaussian density variation (peak density $n_0$ and standard deviation $\sigma$) along the diagnostic line of sight. 

The radiation transport solver is first used to directly solve a single-objective optimization problem, which varies the values of $n_0$, $T_0$, and $\sigma$ to minimize the mean squared deviation between a generated spectrum and the target spectrum. This approach provides reliable fits to the target spectrum, robust to noise and stray line radiation; however, it can be time-intensive, which limits its usefulness for rapid or real-time analysis of large datasets.

We then use the radiation transport solver to generate a $[2500 \times 400]$  dataset of synthetic emission spectra, and we compare the performance of different ML models on their ability to predict density, temperature, and $\sigma$ from the given intensity spectra. The AutoGluon model performs best, with an R2-score of roughly $98\%$ for density and temperature predictions. Simpler models (random forest, k-nearest neighbor, and deep neural network) also exhibit high R2-scores ($>90\%$) for density and temperature predictions, showing the potential of ML models in providing rapid and accurate analysis of spectral data. However, the prediction of $\sigma$ is relatively poor, which is typical of radiation transport problems, and is  related to the relatively smaller sensitivity of the optical depth on this parameter, when compared to the peak density $n_0$.

The mean square deviation from the actual spectra of the ML predictions is typically larger than that of the more time-intensive optimization task. One way to improve the accuracy of these models could be to perform further optimization by including the ML predictions in the initial population; this can provide faster convergence of the optimization and better fitting to the target spectra.


\section*{Acknowledgment}
This work was funded by NSF and NNSA under grant no. PHY2108050, by the EAGER grant no. PHY2213898, and supported by the MathWorks fellowship. 



%

\bibliographystyle{IEEEtran}
\bibliography{main}

\end{document}